\documentclass[aps,pra,onecolumn,superscriptaddress,10pt,nofootinbib]{revtex4-2}
\usepackage{multirow,amsthm,amssymb,amsbsy,amsmath,epsfig,epstopdf,bm,float}
\DeclareMathOperator{\Tr}{Tr}
\usepackage{enumitem}
\usepackage{graphicx}
\usepackage{booktabs}
\usepackage{verbatim}
\usepackage{dcolumn}
\usepackage{color}
\usepackage[dvipsnames]{xcolor}
\usepackage{mathtools}
\usepackage[normalem]{ulem}
\usepackage{soul}
\usepackage{hyperref}
\usepackage{braket}
\usepackage[export]{adjustbox}
\DeclarePairedDelimiter\abs{\lvert}{\rvert}%

\newcommand{\mar}[1]{{\color{black} #1 }}

\begin{document}

\title{A brief journey through collision models for multipartite open quantum dynamics
}

\author{Marco Cattaneo}
\email{marco.cattaneo@helsinki.fi }
\affiliation{QTF Centre of Excellence,  
Department of Physics, University of Helsinki, P.O. Box 43, FI-00014 Helsinki, Finland}
\affiliation{Instituto de F\'{i}sica Interdisciplinar y Sistemas Complejos (IFISC, UIB-CSIC), Campus Universitat de les Illes Balears E-07122, Palma de Mallorca, Spain}
\affiliation{Algorithmiq Ltd, Kanavakatu 3C 00160 Helsinki, Finland}

\author{Gian Luca Giorgi}
\affiliation{Instituto de F\'{i}sica Interdisciplinar y Sistemas Complejos (IFISC, UIB-CSIC), Campus Universitat de les Illes Balears E-07122, Palma de Mallorca, Spain}

\author{Roberta Zambrini}
\affiliation{Instituto de F\'{i}sica Interdisciplinar y Sistemas Complejos (IFISC, UIB-CSIC), Campus Universitat de les Illes Balears E-07122, Palma de Mallorca, Spain}

\author{Sabrina Maniscalco}
\affiliation{QTF Centre of Excellence,  
Department of Physics, University of Helsinki, P.O. Box 43, FI-00014 Helsinki, Finland}
\affiliation{Algorithmiq Ltd, Kanavakatu 3C 00160 Helsinki, Finland}
\affiliation{QTF Centre of Excellence, Department of Applied Physics, School of 
Science, Aalto University, FI-00076 Aalto, Finland}

\date{\today }

\begin{abstract}
The quantum collision models are a useful method to describe the dynamics of an open quantum system by means of repeated interactions between the system and some particles of the environment, which are usually termed ``ancillas''. In this paper, we review the main collision models for the dynamics of multipartite open quantum systems, which are composed of several subsystems. In particular, we are interested in models that are based on elementary collisions between the subsystems and the ancillas, and that simulate global and/or local Markovian master equations in the limit of infinitesimal timestep. After discussing the mathematical details of the derivation of a generic collision-based master equation, we provide the general ideas at the basis of the collision models for multipartite systems, we discuss their strengths and limitations, and we show how they may be simulated on a quantum computer. Moreover, we analyze some properties of a collision model based on entangled ancillas, derive the master equation it generates for small timesteps, and prove that the coefficients of this master equation are subject to a constraint that limits their generality. Finally, we present an example of this collision model with two bosonic ancillas entangled in a two-mode squeezed thermal state.
\end{abstract}
\maketitle


\section{Introduction}
The simplest example of an open quantum system is a single quantum particle (e.g., a qubit) interacting with an external environment, such as a thermal bath. If the system-environment coupling is weak and the autocorrelation functions of the environment decay sufficiently fast, it can be shown that the dynamics of the state of the system is described by a Markovian master equation \cite{breuer2002theory}. The complete characterization of the structure of this master equation is arguably the most celebrated result of the theory of open quantum systems. Specifically, it is known that the most general Markovian master equation of an open quantum system is  the Gorini-Kossakowski-Sudarshan-Lindblad (GKLS) master equation \cite{Gorini1976a,Lindblad1976,Chruscinski2017a}. 
The non-diagonal GKLS master equation for an open system with Hilbert space $\mathcal{H}_S$ is written as (note that throughout the paper we will assume $\hbar=1$):
\begin{equation}
    \label{eqn:GKLSsingleS}
    \frac{d}{dt}\rho_S(t)=\mathcal{L}[\rho_S(t)] = -i[H_S,\rho_S(t)]+\sum_{j,k=1}^{{D_S}^2-1} \gamma_{jk}\left(F_j\rho_S(t)F_k^\dagger-\frac{1}{2}\{\rho_S(t),F_k^\dagger F_j\}\right),
\end{equation}
with ${D_S}\equiv\dim(\mathcal{H}_S)$ and where  $\rho_S(t)$ is the density matrix describing the state of the system at time $t$.
$H_S=H_S^\dagger$ is a self-adjoint operator on $\mathcal{H}_S$, driving the free system dynamics. $\{F_j\}_{j=1}^{{D_S}^2-1}$ with $\Tr[F_jF_k^\dagger]=\delta_{jk}$ are some orthonormal traceless operators that together with the identity form a basis in the space of the bounded operators in $\mathcal{H}_S$. We refer to them as the Gorini-Kossakowski-Sudarshan (GKS) operators. The coefficients $\gamma_{jk}$ are the elements of the positive semi-definite Kossakowski matrix $\bm{\gamma}\geq 0$. The generator of the quantum dynamical semigroup \cite{breuer2002theory} driven by Eq.~\eqref{eqn:GKLSsingleS} is the \textit{Liouvillian} $\mathcal{L}$.

Interestingly, starting from the full microscopic model of system+environment and applying the Born-Markov approximations \cite{breuer2002theory} is not the only way to obtain a GKLS master equation for the state of an open quantum system. For instance, the same dynamics may be recovered through a collision model, that is, through repeated interactions between the system and some external particles, which conceptually belong to the environment, that are called \textit{ancillas}. If the duration of each collision (the \textit{timestep} of the collision model) is infinitesimal, and if a new fresh independent ancilla is prepared at every timestep, then it has been shown that, tracing out the degrees of freedom of the ancillas, we recover a GKLS master equation for the state of the system \cite{Ziman2005,Attal2006a}. 

The interest in quantum collision models has rapidly grown in the past few years. Indeed, this formalism can be successfully applied to different fields where open quantum systems play a key role, including quantum thermodynamics \cite{Barra2015,Strasberg2017,DeChiara2018a} and non-Markovian quantum evolutions \cite{Ciccarello2013a,Vacchini2014,Kretschmer2016a}. A comprehensive review on quantum collision models has been recently published \cite{Ciccarello2021}, as well as a special issue of the journal Entropy \cite{entropySpecial}, a perspective article \cite{Campbell2021a} and a tutorial \cite{Cusumano2022}.

The aim of this work is to explore the main collision models for the dynamics of a multipartite open quantum system, that is, an open system composed of multiple subsystems. The interest in this kind of open quantum systems is broad, owing, for instance, to their relevance for collective dissipative phenomena \cite{Cattaneo2021b}, correlated noise in quantum computers \cite{Sarovar2020}, decoherence-free subspaces \cite{Lidar2003}, and quantum thermodynamics \cite{Levy2014a,Gonzalez2017,Hofer2017}. More specifically, in this paper, we focus on collision models whose description is based on elementary collisions between the subsystems and the ancillas\footnote{This condition makes the collision model straightforwardly implementable on a quantum computer, see for instance the experiments on near-term devices in Refs.~\cite{Garcia-Perez2020,Cattaneo2022}.}, and that can reproduce a (local or global \cite{Levy2014a,Trushechkin2016,Gonzalez2017,Hofer2017,Cattaneo2019b,Scali2020}) GKLS master equation in the limit of infinitesimal timestep. We identify four main schemes of this kind, namely the \textit{multipartite collision model} \cite{Cattaneo2021}, the \textit{cascade collision model} \cite{Giovannetti2012}, the \textit{composite collision model} \cite{Lorenzo2017}, and the \textit{collision model with entangled ancillas} \cite{Daryanoosh2018}. After presenting the derivation of the GKLS master equation for a generic collision model in Section~\ref{sec:collModelSingle}, we will discuss the general idea and the possible limitations of the above-mentioned collision models for multipartite systems in Section~\ref{sec:collModels}. We will also pictorially show how these models may be simulated on a quantum computer. As the properties of the collision model with entangled ancillas are relatively unexplored, we will devote Section~\ref{sec:properties} to a new discussion about them, where we will also analyze the master equation it generates and its limitations. Some concluding remarks are drawn in Section~\ref{sec:conclusions}.

\section{The standard derivation of a collision model for a single open quantum system}
\label{sec:collModelSingle}

In this section, we will provide the standard derivation of a collision model for an individual open quantum system\footnote{Our discussion is mostly based on the derivations in Refs.~\cite{Giovannetti2012,Landi2014,Lorenzo2017}.}. The open system, labeled by ``$S$'', is considered as a single quantum system whose inner structure is not taken into account in the collision model. Our aim is to formalize the open Markovian evolution of $S$ driven by $n$ repeated collisions with different external ancillas labeled by $j=1,\ldots,n$, in addition to a free system dynamics that is independent of the ancillas. 

 We introduce the Hilbert space of the system $\mathcal{H}_S$, whereas the Hilbert space of $j$th ancilla is $\mathcal{H}_{E_j}$. Each ancilla $j$ is initialized in the state $\rho_{E_j}$, and we assume that the state of the system before the $j$th collision is $\rho_S( (j-1)\Delta t)$, where we refer to $\Delta t$ as the \textit{timestep} of the collision model. Then, after the collision, the system state becomes:
\begin{equation}
\label{eqn:collisionJth}
    \rho_{S}(j\Delta t)=\Tr_{E_j}[U_{C,j}(\Delta t)\rho_S((j-1)\Delta t)\otimes\rho_{E_j}U_{C,j}^\dagger(\Delta t)].
\end{equation}
The unitary operator describing the collision is given by:
\begin{equation}
    \label{eqn:collisionInteraction}
    U_{C,j}(\Delta t) = \exp[-{i} \left(g_S H_S+g_I H_{I,j}\right)\Delta t],
\end{equation}
where $H_S$ is the free system Hamiltonian (acting on $\mathcal{H}_S$ only) and $H_{I,j}$ is the interaction Hamiltonian, which for convenience are both dimensionless. The magnitudes of their energies are captured respectively by $g_S$ and $g_I$.

From now on, we will assume that all the ancillas are identical and that they are initialized in the same state $\rho_E$. Moreover, we will employ the same interaction Hamiltonian $H_I$ for all collisions. Then, we can introduce the quantum map $\phi_{\Delta t}$ to represent the system evolution after a single collision between the system and a generic ancilla:
\begin{equation}
    \label{eqn:collisionQuantumMap}
    \phi_{\Delta t}[\boldsymbol{\cdot}] = \Tr_E[U_C(\Delta t) \boldsymbol{\cdot}\otimes\rho_E U_C^\dagger(\Delta t)].
\end{equation}
The map acts on the density matrices of the system alone 
(represented by the dot in the above equation), $E$ refers to the degrees of freedom of a single generic ancilla, while $U_C$ is still given by Eq.~\eqref{eqn:collisionInteraction}, dropping the dependency on $j$. Then, if the state of the system is initialized in $\rho_S(0)$, after $n$ collisions we obtain:
\begin{equation}
\label{eqn:collisionEvn}
    \rho_S(n\Delta t)=\phi_{\Delta t}^n[\rho_S(0)].
\end{equation}

Our goal is to obtain a Markovian master equation from the dynamics driven by the quantum map in Eq.~\eqref{eqn:collisionQuantumMap}. Note that the fact that the ancillas are uncorrelated and collide only once against the system means that the system evolution during the $j$th collision does not depend on the previous history of $\rho_S(t)$, i.e., the Markovianity between the collisions is already enforced by Eq.~\eqref{eqn:collisionEvn}. Then, what we are looking for is a ``smooth'' evolution of $\rho_S(j\Delta t)$ to mimic the GKLS dynamics in Eq.~\eqref{eqn:GKLSsingleS}. Intuitively, we can obtain such a dynamics by assuming that the timestep is infinitesimal, i.e., $\Delta t\rightarrow 0^+$.

Assuming $\Delta t\rightarrow 0^+$, we employ the Baker-Campbell-Hausdorff formula to expand the collision operators in Eq.~\eqref{eqn:collisionQuantumMap}, and we obtain:
\begin{equation}
\label{eqn:collisionDer1}
\begin{split}
    \phi_{\Delta t}[\rho_S]=&\rho_S-{i}\Delta t\Tr_E[[g_S H_S+ g_I H_I,\rho_S\otimes\rho_E]]\\
    &-\frac{\Delta t^2}{2}\Tr_E[[g_S H_S+ g_I H_I,[g_S H_S+ g_I H_I,\rho_S\otimes\rho_E]]]+O(\Delta t^3).
\end{split}
\end{equation}
We now make two crucial assumptions to simplify the above equation. The first one is $\Tr_E[[H_I,\rho_S\otimes\rho_E]]=0$ for all $\rho_S$. This is always valid, for instance, if we decompose the interaction Hamiltonian into system operators $A_\alpha$ and ancilla operators $B_\alpha$, as usually done in the derivation of a Markovian master equation from a microscopic model \cite{breuer2002theory}, i.e.,
\begin{equation}
    \label{eqn:collisionIntDecomp}
H_I=\sum_\alpha A_\alpha\otimes B_\alpha,
\end{equation} 
 and all the ancilla operators are such that $\Tr_E[B_\alpha\rho_E]=0$. The second assumption reads $g_S\ll g_I$ (that is, the collision energy is much larger than the system energy) and $g_I^2\Delta t\rightarrow \gamma$, $g_S^2\Delta t \rightarrow 0$ for $\Delta t\rightarrow 0^+$, where $\gamma$ is a finite constant with the units of the inverse of time. Under these assumptions, Eq.~\eqref{eqn:collisionDer1} becomes:
\begin{equation}
\label{eqn:collisionDer2}
\begin{split}
    \phi_{\Delta t}[\rho_S]=&\rho_S-{i}\Delta t g_S [H_S,\rho_S]
    +\gamma\Delta t \Tr_E[H_I\rho_S\otimes\rho_E H_I-\frac{1}{2}\{H_I^2,\rho_S\otimes\rho_E\}]\\&+O(g_S^2\Delta t^2)+O(g_Sg_I\Delta t^2)+O(\Delta t^3).
\end{split}    
\end{equation}

For a single collision, $\rho_S(t+\Delta t)=\phi_{\Delta t}[\rho_S(t)]$, so: 
\begin{equation}
\label{eqn:collisionGKLS}
\begin{split}
    \frac{d}{dt}\rho_S(t)=&\lim_{\Delta t\rightarrow 0^+}\frac{\rho_S(t+\Delta t)-\rho_S(t)}{\Delta t}=\lim_{\Delta t\rightarrow 0^+}\frac{\phi_{\Delta t}[\rho_S(t)]-\rho_S(t)}{\Delta t}\\
    =&-{i} g_S [H_S,\rho_S(t)]+\gamma\sum_{\alpha,\alpha'}\Tr_E[B_\alpha^\dagger B_{\alpha'}\rho_E]\Big(A_{\alpha'}\rho_S(t) A_{\alpha}^\dagger
     -\frac{1}{2}\{A_\alpha^\dagger A_{\alpha'},\rho_S(t)\}\Big),
\end{split}
\end{equation}
where we have used Eqs.~\eqref{eqn:collisionIntDecomp} and~\eqref{eqn:collisionDer2}, and the assumptions on the magnitude of $g_S$ and $g_I$. The above equation is in the non-diagonal GKLS form with GKS operators given by $\{A_\alpha\}$. Indeed, $H_S^\dagger=H_S$, and the correlation matrix $\Tr_E[B_\alpha^\dagger B_{\alpha'}\rho_E]$ is semi-positive definite\footnote{Indeed, let us introduce the matrix $M_{\alpha\alpha'}=\Tr_E[B_\alpha^\dagger B_{\alpha'}\rho_E]$. Then, note that, given any complex vector $\mathbf{v}$, we have $\sum_{\alpha\alpha'}v_\alpha^* M_{\alpha\alpha'}v_{\alpha'}=\Tr_E[C^\dagger C \rho_E]\geq 0$, because the state of the ancilla is positive semi-definite. We have introduced the operator $C=\sum_\alpha v_\alpha B_\alpha$.}. Therefore, we can reproduce the action of a quantum dynamical semigroup driven by the Liouvillian $\mathcal{L}$ at any time $t$ by performing $n$ repeated collisions between system and ancillas, with $n=t/\Delta t$. That is to say:
\begin{equation}
    \label{eqn:collisionDynSemig}
    \exp(\mathcal{L}t)=\lim_{\Delta t\rightarrow 0^+}\phi_{\Delta t}^n, \text{ with $n=t/\Delta t$}.
\end{equation}

Finally, note that, assuming the same approximations, we obtain the same result and the same master equation if we decompose the collision operator as follows:
\begin{equation}
\label{eqn:opDecomposedSI}
    U_C(\Delta t)\rightarrow \underbrace{e^{-{i}g_S H_S \Delta t}}_{U_I(\Delta t)}\underbrace{e^{-{i}g_I H_I \Delta t}}_{U_S(\Delta t)},
\end{equation}
or equivalently $U_C(\Delta t)\rightarrow U_S(\Delta t) U_I(\Delta t)$. Indeed, the error due to the non-commutativity of $H_S$ and $H_I$ is of the order of $g_S g_I \Delta t^2$, and therefore it is absorbed by the corresponding remainder in Eq.~\eqref{eqn:collisionDer2}.
This means that we can first perform the collision and then, independently, the free system evolution (or vice versa). This result allows us to tune $g_I$ and $g_S$ by using different effective collision timesteps.

\section{Collision models for multipartite open quantum systems}
\label{sec:collModels}
In this section, we review the main collision models for multipartite open quantum systems that have been introduced to date. Our goal is to understand the crucial idea at the basis of each of them, their major limitations, and which situations they are suitable for.  Moreover, to stress the connection between the formalism of collision models and ancilla-based quantum simulation algorithms, we will depict the elementary mechanism of each collision model through gates applied on the system and ancillary qubits of a quantum computer. We will not show here the mathematical details of the derivations of these collision models\footnote{With the exception of the collision model with entangled ancillas, for which we provide a new formalization in Section~\ref{sec:properties}.}, and we suggest  checking the original papers for further details. Anyway, to get a grasp on these mathematical details, the reader may get familiar with the derivation of the collision model for a single open system we presented in Section~\ref{sec:collModelSingle}, which is at the basis of all the collision schemes we will discuss.

We will write here the expression of the GKLS master equation for a multipartite open quantum system we aim to simulate through a collision model. Let us suppose that the open system is composed of $M$ (for simplicity identical) subsystems. Its Hilbert space can be written as $\mathcal{H}_S=\bigotimes_{m=1}^M\mathcal{H}_{S_m}$. The dimension of each subsystem is $\dim(\mathcal{H}_{S_m})=d$. Then, $D_S=\dim(\mathcal{H}_S)=d^M$. For simplicity, we will focus on multipartite GKLS equations that can be structured such that their GKS operators $F_k$ in Eq.~\eqref{eqn:GKLSsingleS} act locally on a single subsystem only. Anyway, extensions to many-body GKS operators can be  made for all the collision models we will analyze. Under the assumption of local GKS operators, we can write the multipartite Markovian master equation as:
\begin{equation}
\label{eqn:gklsMultipartite}
    \frac{d}{dt}\rho_S(t)=\mathcal{L}[\rho_S(t)]=-i[H_S,\rho_S(t)]+\sum_{m,\alpha,m',\alpha'}\gamma_{m,\alpha,m',\alpha'}\left(F_{m,\alpha}\rho_S(t) F_{m',\alpha'}^\dagger-\frac{1}{2}\{ F_{m',\alpha'}^\dagger F_{m,\alpha},\rho_S(t)\}\right).
\end{equation}
$H_S$ is the effective Hamiltonian driving the free system dynamics. The indexes $m,m'=1,\ldots,M$ run over the subsystems, while $\alpha,\alpha'$ run over the different local GKS operators of the open dynamics. If the Hilbert space of each subsystem has dimension $d$, then $\alpha,\alpha'=1,\ldots,d^2-1$, according to the discussion after Eq.~\eqref{eqn:GKLSsingleS}. $F_{m,\alpha}$ is a GKS operator that is local on the subsystem $m$, while $\gamma_{m,\alpha,m',\alpha'}$ are the coefficients of the master equation\footnote{The reader should be aware that, for convenience, in  Eq.~\eqref{eqn:gklsMultipartite} we are not expressing the fact that, because the GKLS master equation generates a completely-positive-trace-preserving map \cite{breuer2002theory}, for each term in the master equation with, for instance, $F_{m,\alpha}\rho_S F_{m',\alpha'}^\dagger$, there must be a term $F_{m',\alpha'}\rho_S F_{m,\alpha}^\dagger$ with a complex conjugate coefficient. We will label both these terms through the same quartet $(m,\alpha,m',\alpha')$.}.

We point out that we will not discuss here the collision models for multipartite systems put forward in different papers by \c{C}akmak et al. to study entanglement generation \cite{Cakmak2019}, quantum synchronization \cite{Karpat2019,Karpat2021} and quantum Darwinism \cite{Cakmak2021}, as it is not clear how to derive a GKLS master equation for these models (the authors do not consider the limit of infinitesimal timestep). However, these schemes are useful to study open multipartite dynamics via the formalism of quantum maps with a finite timestep, and we refer the readers to the original papers for further details. Moreover, a collision scheme leading to multipartite open dynamics driven by a GKLS master equation has been recently employed to study the charging of a quantum battery \cite{PhysRevA.105.062203}. This collision model, which we will not discuss in this section, creates cross terms between the subsystems by engineering a many-body non-linear collision Hamiltonian that couples both the subsystems and the ancillas at the same time. \mar{In a similar spirit, a collective subsystems-ancilla interaction is engineered in recent collision models to study non-stationary phenomena of open spin systems \cite{Li2022} or to explore quantum thermodynamics with common reservoirs \cite{Huang2022}. Finally, two recent papers \cite{Gillman2022,Gillman2022a} have shown that a quantum version of the cellular automata may implement the closed and open quantum dynamics of many-body systems. This formalism makes use of repeated gates that propagate the state of the system in time, and it is therefore analogous to quantum collision models.}

\subsection{Multipartite Collision Model (MCM)}
\label{sec:MCM}
Together with Gabriele De Chiara, we have recently introduced the \textit{Multipartite Collision Model} (MCM), which is able to reproduce the most general Markovian dynamics of any open quantum system \cite{Cattaneo2021}. For this reason, the MCM may be considered the most general and complete scheme among the ones we are presenting here. The price to pay for such a generality lies in the fact that the MCM requires one ancillary qubit for each unordered pair of jump operators $F_{m,\alpha}$ and $F^\dagger_{m',\alpha'}$ in Eq.~\eqref{eqn:gklsMultipartite}, i.e., for each quartet $(m,\alpha,m',\alpha')$. However, several shortcuts may be engineered depending on the symmetries of the multipartite system and of the master equation, and the number of necessary ancillas may be actually much lower. We refer the readers to the supplementary material of the original paper for further details on these shortcuts \cite{Cattaneo2021}.

\begin{figure}
    \centering
    \includegraphics[scale=0.16]{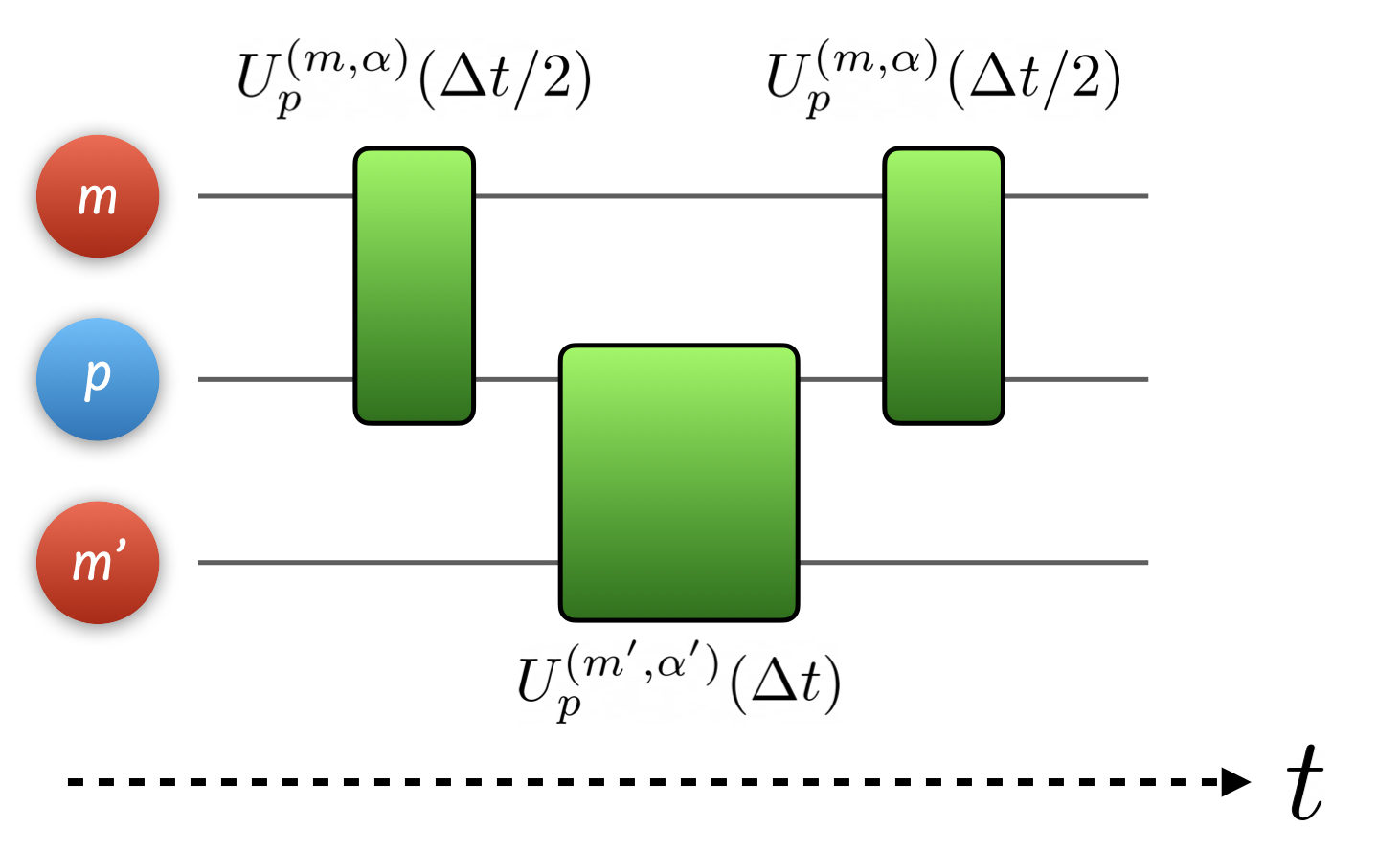}
    \caption{Structure of the collisions of a single ancilla $p$ (blue qubit) of the MCM with the subsystems $m$ and $m'$ (red qubits) implemented on a quantum computer. Each green interaction represents a two-qubit gate, which plays the role of an elementary collision.}
    \label{fig:MCM}
\end{figure}

The action of a single ancilla of the MCM is depicted in Fig.~\ref{fig:MCM}, assuming, for simplicity, that the subsystems of the multipartite system are qubits. The MCM reproduces each single ``brick'' of the GKLS master equation, that is, each term coupling the jump operators $F_{m,\alpha}$ and $F^\dagger_{m',\alpha'}$ in Eq.~\eqref{eqn:gklsMultipartite} (including the ones with $m=m'$ and $\alpha=\alpha'$), through an interaction $U_p$ between the subsystems $m,m'$ and an ancilla labeled by $p$, with $p=(m,\alpha,m',\alpha')$, which can be decomposed into three elementary collisions via the following formula:
\begin{equation}
\label{eqn:intOp}
    U_p(\Delta t)=U_p^{(m,\alpha)}(\Delta t/2)U_p^{(m',\alpha')}(\Delta t)U_p^{(m,\alpha)}(\Delta t/2).
\end{equation}
Each elementary collision is given by:
\begin{equation}
\label{eqn:ElColMCM}
\begin{split}
    U_p^{(m,\alpha)}(\Delta t)&=\exp(-i g_I\Delta t (\lambda_{m,\alpha}F_{m,\alpha}\otimes\sigma_{E,p}^++H.c.)),
\end{split}
\end{equation}
where $\sigma_{E,p}^+$ is the operator that creates an excitation in the ancillary qubit, while $\lambda_{m,\alpha}$  are dimensionless parameters that we can freely tune. We notice that the GKS operators are ``encoded'' in the interaction Hamiltonian of each elementary subsystem-ancilla collision.

The interaction operator $U_p(\Delta t)$ reproduces the corresponding term in the master equation by taking the standard limits $\Delta t\rightarrow 0^+$ and $g_I^2\Delta t\rightarrow \gamma$, according to the discussion in Section~\ref{sec:collModelSingle}. The key idea behind this result lies in the structure of the decomposed ancilla-subsystems interaction, which is based on the \textit{second-order} Suzuki-Trotter formula \cite{Hatano2005}:
\begin{equation}
    \label{eqn:secondorderSuzuki}
    e^{\frac{\Delta t}{2}A}e^{\Delta t B}e^{\frac{\Delta t}{2}A}= e^{\Delta t(A+B)}+O(\Delta t^3),
\end{equation}
for any bounded operators $A,B$ on the Hilbert space of the system.
The reader can verify that inserting $U_p(\Delta t)$ in Eq.~\eqref{eqn:collisionQuantumMap} (or analogously in Eq.~\ref{eqn:collisionDer1}), and then using Eq.~\eqref{eqn:secondorderSuzuki}, yields the correct result. \mar{Note that, clearly, switching $A$ and $B$ does not change the result, and the same holds for switching $m$ and $m'$ in Eq.~\eqref{eqn:intOp}.}

We can build an interaction operator as in Eq.~\eqref{eqn:intOp} for all possible ancillas $p=(m,\alpha,m',\alpha')$. Then, their composition in any order will produce the most general dissipator of  Eq.~\eqref{eqn:gklsMultipartite}, if the ancillas are initialized in the ground state. Moreover, the free system dynamics driven by $H_S$ can be added following the procedure discussed in Section~\ref{sec:collModelSingle}, and the most general GKLS equation for a multipartite system is simulated in the limit of infinitesimal timestep. That is, the quantum map for the MCM reproduces any quantum dynamical semigroup driven by a generic $\mathcal{L}$, the latter given by Eq.~\eqref{eqn:gklsMultipartite}, according to Eq.~\eqref{eqn:collisionDynSemig}.

Ref.~\cite{Cattaneo2021} shows that the MCM is efficiently simulable on a quantum computer under standard assumptions, and it also discusses how to extend this formalism to non-local GKS operators and how to compute the error due to a small but finite $\Delta t$. Finally, the experimental implementation of the multipartite collision model on a near-term quantum computer has been recently presented \cite{Cattaneo2022}.

\subsection{Cascade collision model}
\label{sec:cascade}
\begin{figure}
    \centering
    \includegraphics[scale=0.16]{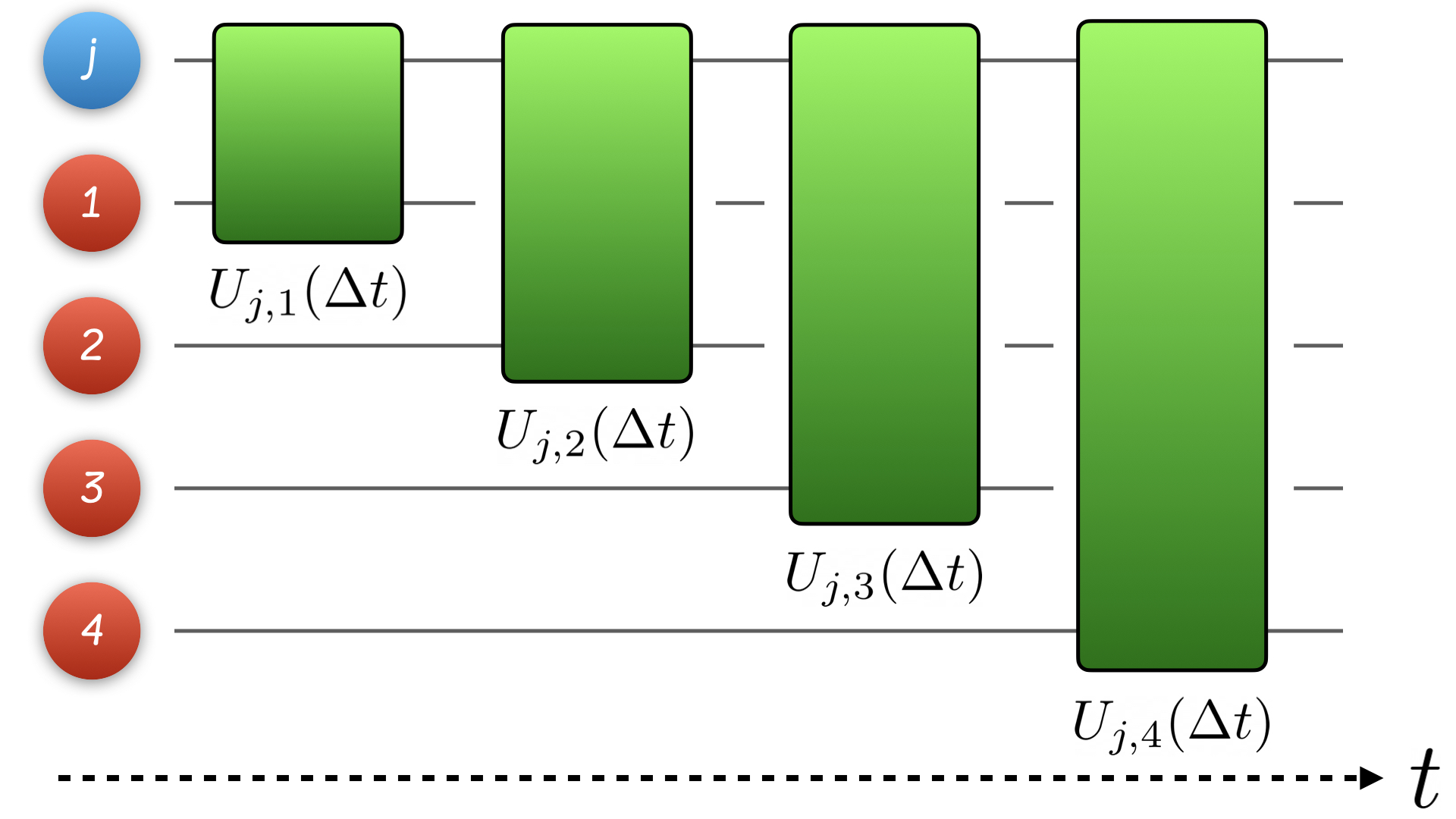}
    \caption{Structure of a single timestep of the cascade collision model implemented on a quantum computer with $M=4$. The ancilla $j$ (blue qubit) collides with the subsystem qubit 1, then with qubit 2, and so on (red qubits). Each green interaction represents a two-qubit gate.}
    \label{fig:cascade}
\end{figure}

To the best of our knowledge, the first collision model for multipartite open quantum systems was introduced by Giovannetti and Palma in 2012 \cite{Giovannetti2012,Giovannetti2012a}. This collision model reproduces a \textit{cascade interaction}, that is, a single ancilla collides with the first subsystem of the open quantum system, then it goes away and collides with the second subsystem, then with the third one, and so and so forth. The interaction of the ancilla with all the subsystems (from $m=1$ to $m=M$) represents a single timestep $\Delta t$ of the cascade collision model. Additional timesteps may then be implemented by repeating the same collision pattern. This scheme is depicted in Fig.~\ref{fig:cascade} for a single ancillary qubit (hence, for a single timestep) and four system qubits ($M=4$).

Following the original papers, we can encode the jump operators of the master equation we aim to reproduce in the collision Hamiltonian between the ancilla and every single subsystem. Specifically, the operator describing the interaction between the $j$th ancilla and the $m$th subsystem in the cascade collision model is \cite{Giovannetti2012,Giovannetti2012a}:
\begin{equation}
    \label{eqn:cascadeInt}
    U_{j,m}(\Delta t)=\exp(-i g_I H_{j,m} \Delta t),
\end{equation}
with
\begin{equation}
    \label{eqn:cascadeHam}
    H_{j,m}=\sum_{\alpha} F_{m,\alpha}\otimes B_{j,\alpha}, 
\end{equation}
where $\{B_{j,\alpha}\}$ is a family of ancilla operators that satisfy the standard assumption (e.g., see the discussion related to Eq.~\eqref{eqn:collisionIntDecomp}) $\Tr_{E_j}[B_{j,\alpha}\rho_{E_j}]=0$, while $\rho_{E_j}$ is the initial state of the ancilla. This is valid, for instance, if the ancilla is a qubit initialized in the ground state and all the $B_{j,\alpha}$ are $\sigma_{E_j}^+$, as for the multipartite collision model. The free system dynamics driven by $H_S$ can then be inserted following the procedure discussed in Section~\ref{sec:collModelSingle} \cite{Giovannetti2012a}. It can be shown \cite{Giovannetti2012,Giovannetti2012a} that this collision model in the standard limits of infinitesimal timestep and $g_I^2\Delta t\rightarrow \gamma$ leads to a master equation that can be written as follows:
\begin{equation}
\label{eqn:cascadeGKLS}
    \frac{d}{dt}\rho_S(t)=-i[H_S,\rho_S(t)]+\sum_{m=1}^M \mathcal{D}^{(loc)}_m[\rho_S(t)]+\sum_{\substack{m,m'=1\\m'>m }}^M \mathcal{D}^{(glob)}_{m,m'}[\rho_S(t)].
\end{equation}
We observe a GKLS dissipator that acts locally on subsystem $m$
\begin{equation}
    \mathcal{D}^{(loc)}_m[\rho_S]=\sum_{\alpha,\alpha'}\tilde{\gamma}_{m,\alpha,m,\alpha'}\left(F_{m,\alpha}\rho_S F_{m,\alpha'}^\dagger-\frac{1}{2}\{ F_{m,\alpha'}^\dagger F_{m,\alpha},\rho_S)\}\right),
\end{equation}
and a global dissipator that has the following structure:
\begin{equation}
\label{eqn:globalDissipator1}
    \mathcal{D}^{(glob)}_{m,m'}[\rho_S]=\sum_{\alpha,\alpha'}\left(\tilde{\gamma}_{m,\alpha,m',\alpha'} F_{m,\alpha}[\rho_S,F_{m',\alpha'}^\dagger]-\tilde{\gamma}_{m,\alpha,m',\alpha'}^* [\rho_S,F_{m',\alpha'}]F_{m,\alpha}^\dagger\right).
\end{equation}
\mar{By adding and subtracting the terms $\frac{1}{2}\rho_S F_{m,\alpha}F_{m',\alpha'}^\dagger $ and $\frac{1}{2}F_{m',\alpha'}F_{m,\alpha}^\dagger\rho_S$, we can write $\mathcal{D}^{(glob)}_{m,m'}$ in a manifestly GKLS form \cite{Cusumano2017a}:
\begin{equation}
\label{eqn:globalDissipator2}
\begin{split}
    \mathcal{D}^{(glob)}_{m,m'}[\rho_S]=&-i \left[H_{LS}^{(m,m')},\rho_S\right]+\sum_{\alpha,\alpha'}\tilde{\gamma}_{m,\alpha,m',\alpha'} \left(F_{m,\alpha}\rho_S F_{m',\alpha'}^\dagger-\frac{1}{2}\{F_{m,\alpha}F_{m',\alpha'}^\dagger,\rho_S\}\right)\\
    &+\sum_{\alpha,\alpha'}\tilde{\gamma}_{m,\alpha,m',\alpha'}^* \left(F_{m',\alpha'}\rho_S F_{m,\alpha}^\dagger-\frac{1}{2}\{F_{m',\alpha'}F_{m,\alpha}^\dagger,\rho_S\}\right),
\end{split}
\end{equation}
where we have introduced the Hermitian operator $H_{LS}^{(m,m')}$ that plays the role of an effective Lamb-shift Hamiltonian:
\begin{equation}
\label{eqn:lambShift}
H_{LS}^{(m,m')}=\sum_{\alpha,\alpha'}\frac{\tilde{\gamma}_{m,\alpha,m',\alpha'}F_{m,\alpha}F_{m',\alpha'}^\dagger- \tilde{\gamma}_{m,\alpha,m',\alpha'}^* F_{m',\alpha'}F_{m,\alpha}^\dagger }{2i}.
\end{equation}
}

In the original papers, the coefficients $\tilde{\gamma}_{m,\alpha,m',\alpha'}$ are tuned by inserting a general quantum map, which acts locally on the state of the ancilla only, between two subsystem-ancilla collisions. The properties of this quantum map modify the autocorrelation functions $\Tr_{E_j}[B_{j,\alpha'}^\dagger B_{j,\alpha} \rho_{E_j}]$ on which the coefficients depend. This cannot be straightforwardly represented in Fig.~\ref{fig:cascade}, as reproducing a non-unitary dynamics on a quantum computer is not a trivial procedure and typically requires some additional ancillary qubits \cite{Lloyd2001,Koniorczyk2006,Barreiro2011,Cleve2017,Cattaneo2021}. Alternatively, one may freely tune these coefficients by implementing more collisions whose interaction Hamiltonian in Eq.~\eqref{eqn:cascadeHam} depends on controllable parameters, as for the MCM.

The form of the cascade master equation in Eq.~\eqref{eqn:cascadeGKLS} is not as general as Eq.~\eqref{eqn:gklsMultipartite} because of the condition $m'>m$ in the global dissipator. This structure is indeed capturing the ``causality'' of the cascade interaction, i.e., the fact that the ancilla is colliding with some qubits before than with some others. \mar{This can be seen by performing the partial trace over all the subsystems with $m>1$ on Eq.~\eqref{eqn:cascadeGKLS}, which would yield a Markovian local dynamics for the subsystem 1 only \cite{Giovannetti2012}. Curiously, this causal order is encoded in the Lamb-shift Hamiltonians defined by Eq.~\eqref{eqn:lambShift}. Without these terms, the master equation would be equivalent to the one generated by the MCM. This means that, to obtain the master equation simulated by the MCM, one may also implement the cascade collision model and then simulate a free system dynamics driven by a counter-Hamiltonian that cancels out the Lamb-shift terms in Eq.~\eqref{eqn:lambShift}. We clarify this point through an example in Appendix~\ref{sec:exDiff}.}

To obtain the most general GKLS master equation for multipartite systems starting from Eq.~\eqref{eqn:cascadeGKLS}, one may implement additional cascaded collisions in which the ancilla first interacts with the last qubit, and so on until the first qubit (i.e., we reverse the order of the collisions, and the requirement in Eq.~\eqref{eqn:cascadeGKLS} becomes $m'<m$). To be really general, one should implement any possible permutation of the order of the subsystems in the global dissipator. Such a scheme, however, resembles the structure of the interaction of the multipartite collision model in Eq.~\eqref{eqn:intOp}, which is already optimal due to the second-order Suzuki-Trotter decomposition. Therefore, the MCM can be thought of as the optimal extension of the cascade collision model to the most general multipartite Markovian dynamics, where there is no ``preferred time order'' of the collisions between the ancillas and the subsystems. \mar{Intuitively, the two ``spatial'' directions of the MCM elementary collisions in Eq.~\eqref{eqn:intOp} and Fig.~\ref{fig:MCM} (first with subsystem $m$, then with subsystem $m'$, and then again with subsystem $m$) are creating and then destroying the evolution generated by the Lamb-shift Hamiltonian in Eq.~\eqref{eqn:lambShift}, thus removing any causal order. To have a better grasp on this issue, we refer the readers to the example in Appendix~\ref{sec:exDiff}.} 

The cascade open quantum dynamics has been proven useful to study, for instance, Landauer's principle in multipartite systems \cite{Lorenzo2015a}, the heat flux and quantum correlations produced by the cascade interaction \cite{Lorenzo2015}, interferometry \cite{Cusumano2017a,Cusumano2018},  giant emitters in quantum optics \cite{Carollo2020,Cilluffo2020}, and steady-state thermodynamics \cite{li2022steady}. \mar{Interestingly, in 1991 Bergou and Hillery presented a study on a model of two masers pumped by the same beam of excited atoms that can be formalized exactly as a cascade collision model \cite{Bergou1991}. The master equation for the state of the two masers has the same structure as Eq.~\eqref{eqn:cascadeGKLS}.}

\subsection{Composite collision model for local master equations}
The composite collision model was introduced by Lorenzo, Ciccarello and Palma in 2017 \cite{Lorenzo2017}. Although this collision model is applied to multipartite open quantum systems, the dynamics it can generate corresponds to a local master equation \cite{Levy2014a,Trushechkin2016,Gonzalez2017,Hofer2017,Cattaneo2019b,Scali2020}, that is, it is not able to describe global master equations \cite{Gonzalez2017,Hofer2017,Cattaneo2019b} that can capture collective dissipative phenomena \cite{Cattaneo2021b}. 

\begin{figure}
    \centering
    \includegraphics[scale=0.16]{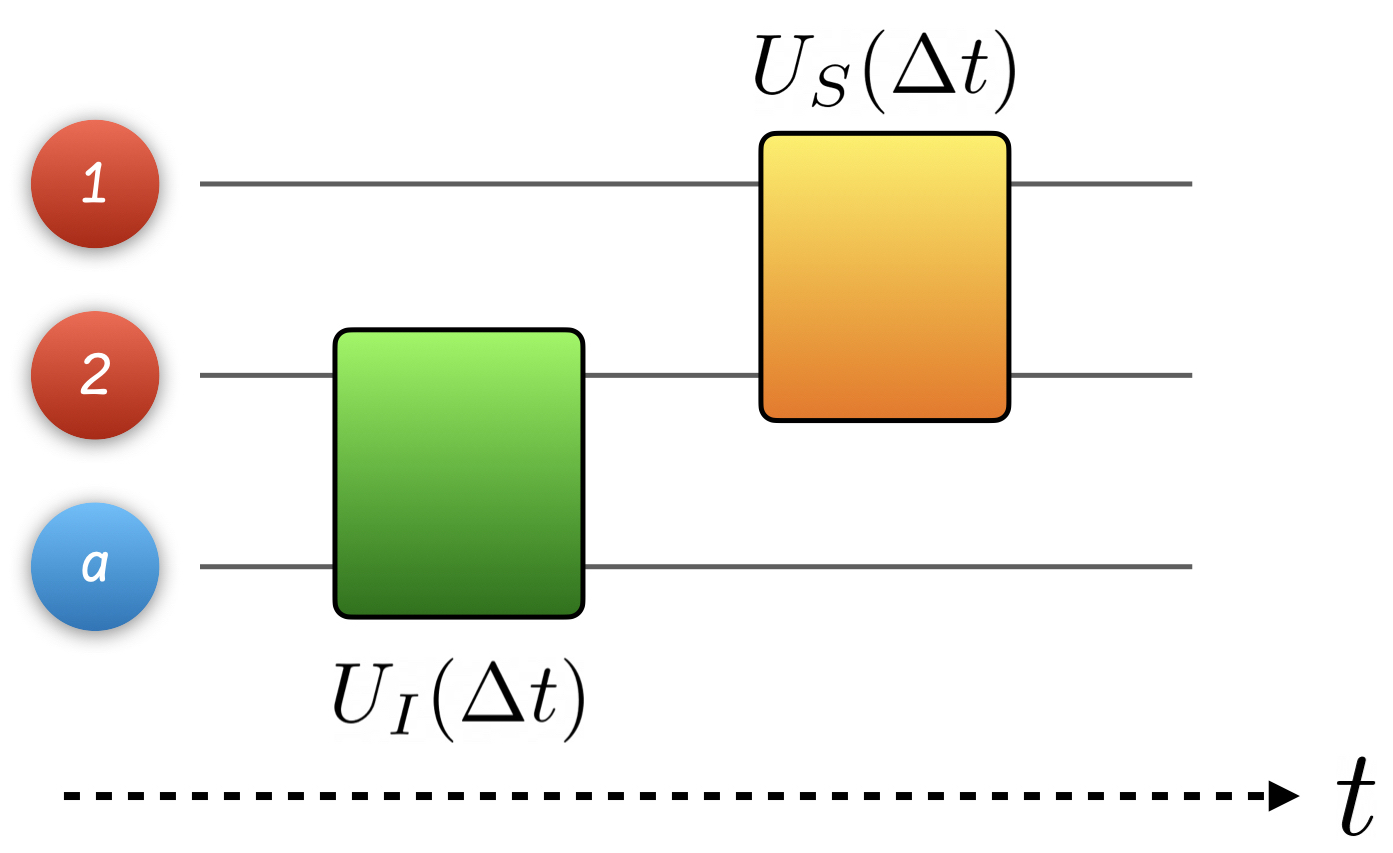}
    \caption{Structure of a single timestep of the composite collision model implemented on a quantum computer with $M=2$. First, the ancilla $a$ (blue qubit) collides with the subsystem qubit 2 through the operator $U_I(\Delta t)$. Next, the two subsystems (red qubits) interact through the free system evolution $U_S(\Delta t)$. The green and orange interactions represent two-qubit gates.}
    \label{fig:composite}
\end{figure}

The idea at the basis of the composite collision model is quite simple but elegant\footnote{Note that an equivalent formulation of this collision model, although based on a different language and with other purposes, can be found in Ref.~\cite{Landi2014}.}. We consider, for simplicity, a system of $M=2$ subsystems. The subsystem $m=1$ interacts only with the second subsystem, while it is not directly coupled to an external environment (i.e., it does not collide with any ancilla). In contrast, the subsystem $m=2$ collides with a fresh ancilla at each timestep $\Delta t$, and the magnitude of this interaction is given by $g_I$. Moreover, a free system dynamics driven by a non-local Hamiltonian $H_S$ with magnitude $g_S$ that couples the two subsystems is implemented between two consecutive collisions (i.e., one free system evolution for each timestep). Therefore, the total evolution for a single timestep can be described by the operator in Eq.~\eqref{eqn:opDecomposedSI}. A single timestep of the composite collision model is depicted in Fig.~\ref{fig:composite}, where both the subsystems and the ancilla are qubits. Note that, even if the first subsystem is not directly interacting with the ancilla, its dynamics is anyway non-unitary because of the interaction with the second subsystem. For instance, the state of the first subsystem can thermalize. 

In the framework of the composite collision model, we can recover a GKLS master equation in the standard limits $\Delta t\rightarrow 0^+$, $g_I^2\Delta t\rightarrow \gamma$, $g_S\ll g_I$, as discussed in Section~\ref{sec:collModelSingle}. This master equation is \textit{local}, that is, its dissipator can be decomposed as a sum of local terms whose jump operators act on a single subsystem only. By considering a generic number of subsystems and implementing all the possible local ancillary collisions in order to tune the coefficients at our will, the most general master equation generated by the composite collision model is:
\begin{equation}
\label{eqn:compositeGKLS}
    \mathcal{L}[\rho_S(t)]=-i[H_S,\rho_S(t)]+\sum_{m,\alpha,\alpha'}\gamma_{m,\alpha,m,\alpha'}\left(F_{m,\alpha}\rho_S(t) F_{m,\alpha'}^\dagger-\frac{1}{2}\{ F_{m,\alpha'}^\dagger F_{m,\alpha},\rho_S(t)\}\right),
\end{equation}
which is equivalent to Eq.~\eqref{eqn:gklsMultipartite} with $m'=m$.

While the composite collision model is able to reproduce the most general \textit{local} Markovian master equation, it was originally introduced to study non-Markovian dynamics by considering as the ``open system'' all the subsystems that are not directly colliding with the ancillas (e.g., subsystem $m=1$ in Fig.~\ref{fig:composite}), while the subsystems that are interacting with the environment particles (e.g., the second subsystem in Fig.~\ref{fig:composite}) are considered as ``pseudomodes''. Tracing out the degrees of freedom of both the ancillas and the pseudomodes, we recover a non-Markovian dynamics for the open system only. This way to describe non-Markovianity has been proven to be particularly successful \cite{Campbell2018,Luchnikov2020}. Furthermore, the formalism of the composite collision model has been employed to study local master equations, mainly for quantum thermodynamic analyses \cite{DeChiara2018a,Hewgill2020,Hewgill2020a,pedram2022environment}, and quantum correlations in multipartite open dynamics \cite{Li2020d}. Moreover, Ref.~\cite{Tian2021} compares the heat currents obtained either through the composite collision model with two system qubits or through a similar model with a collective term in the system Hamiltonian (these two models are the collisional analogue of the local and global master equations \cite{Hofer2017,Gonzalez2017,Cattaneo2019b}), although the infinitesimal timestep limit and the corresponding GKLS master equation are not studied in this work. \mar{Finally, a similar collision model for local master equations (with possible extensions to global ones) can be found in Ref.~\cite{Arisoy2019}. In this work, the derivation of the master equation relies on the standard microscopic theory of open quantum systems \cite{breuer2002theory} and not on the formalism described in Section~\ref{sec:collModelSingle}.}

\subsection{Collision model with entangled ancillas}
\label{sec:collEntAnc}
\begin{figure}
    \centering
    \includegraphics[scale=0.16]{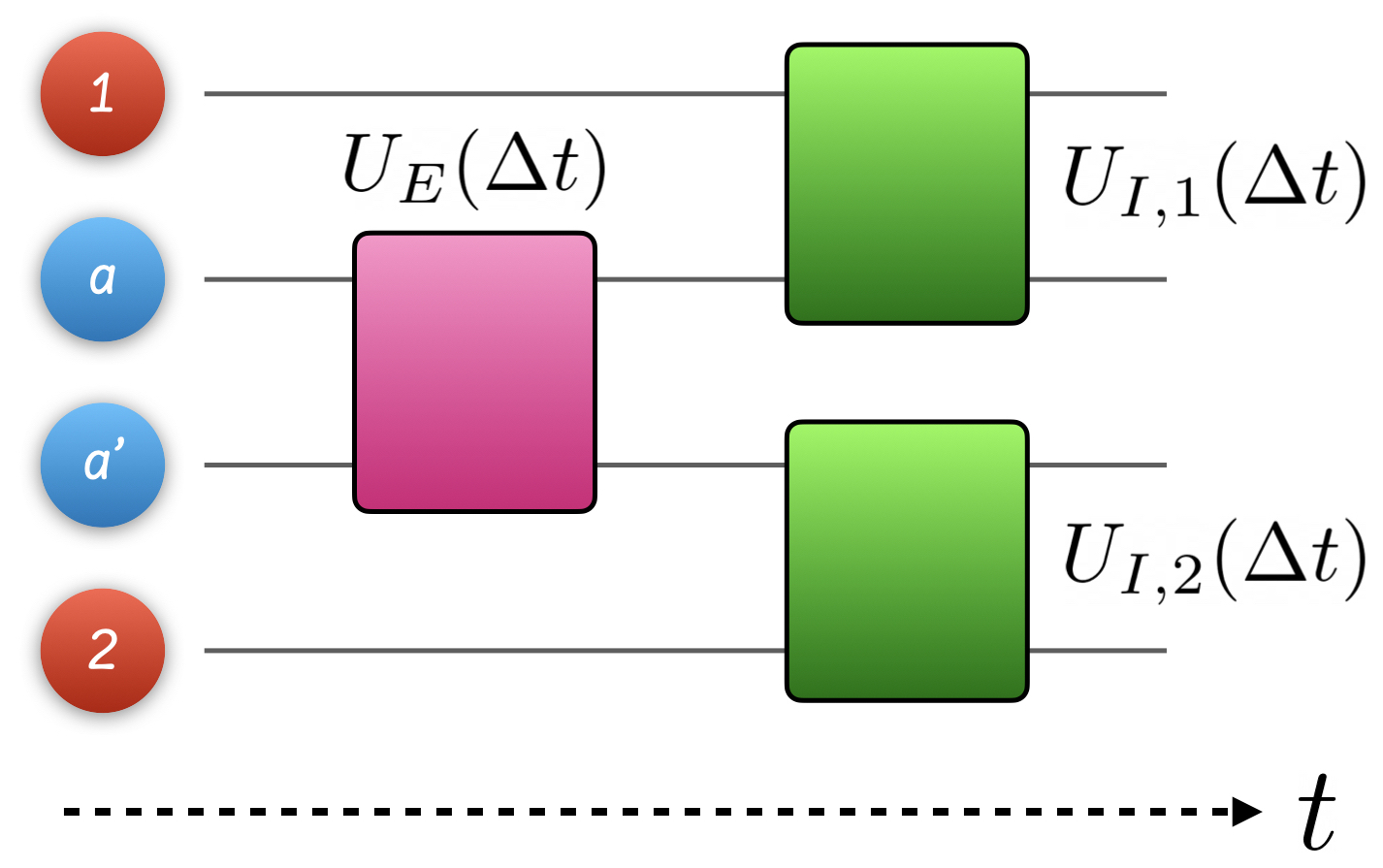}
    \caption{Structure of a single timestep of the  collision model with entangled ancillas implemented on a quantum computer with $M=2$. First, the ancillas $a$ and $a'$ (blue qubits) get entangled through the operator $U_E(\Delta t)$. Next, the ancilla $a$ ($a'$) collides with the subsystem 1 (2) (red qubits) through the operator $U_{I,1}(\Delta t)$ ($U_{I,2}(\Delta t)$). The green and pink interactions represent two-qubit gates.}
    \label{fig:entangled}
\end{figure}

In this subsection, we will introduce the last type of collision model for multipartite open dynamics we consider in this work. Specifically, we will present a collision model that makes use of entangled ancillas at each timestep. Note that this is different from introducing quantum and/or classical correlations between ancillas prepared for the collisions at different timesteps, which is one of the ways to study non-Markovianity in collision models (see Ref.~\cite{Ciccarello2021} and references therein). In contrast, here we are interested in collision models where pairs of entangled ancillas collide at the same time with different subsystems of a multipartite open quantum system. A single timestep of this collision model is depicted in Fig.~\ref{fig:entangled}, where for simplicity we consider a two-qubit open system colliding with two entangled ancillary qubits. To the best of our knowledge, this collision scheme has been introduced first by Daryanoosh et al. in 2018 \cite{Daryanoosh2018}, and then it has been employed to study the thermodynamics of thermal machines powered by correlated baths \cite{DeChiara2020},  the quantum Rayleigh problem \cite{PhysRevResearch.3.023235} and quantum trajectories for correlated ancillas \cite{Daryanoosh2022,Gross2022}.

The properties of the collision model with entangled ancillas have not been studied as much as the ones of the other collision models we have introduced in this section. Specifically, Refs.~\cite{Daryanoosh2018,Daryanoosh2022} derive a master equation for a two-qubit environment whose ancillas are prepared in a linear composition of different Bell states and make some comments about the possible extensions to $m$ qubits. It is our aim to characterize the collision model with entangled ancillas in the language of multipartite open quantum dynamics, so as to recover a GKLS master equation whose structure resembles Eq.~\eqref{eqn:gklsMultipartite}. Therefore, we devote the following Section~\ref{sec:properties} to a discussion on some properties of this collision model.

\section{Properties of the collision model with entangled ancillas}
\label{sec:properties}
In this section we will analyze some properties of the collision model with entangled ancillas that has been introduced in Section~\ref{sec:collEntAnc}. 

\subsection{Model formalization}
First, let us formalize the elementary timestep of this collision model that is shown in Fig.~\ref{fig:entangled} for a simplified scenario. In general, we assume to have $M$ subsystems and $M$ ancillas that will collide locally with each of them. At time $t=0$, the state of the open system is initialized in $\rho_S$, while, to keep the analogy with the ancillary qubits of a quantum computer, we msy assume that the ancillas are all prepared in the ground state. Hence, we can write the total state of the ancillas as $\rho_E(0)=\bigotimes_{j=1}^M \ket{0}_j\!\bra{0}$. Then, a single timestep of the collision model may be described as:
\begin{equation}
    \label{eqn:singleTimeEnt}
    \phi_{\Delta t}^{(ent)}[\rho_S]=\Tr_E[U_{tot}(\Delta t)\rho_S\otimes\rho_E(0)U_{tot}^\dagger(\Delta t)],
\end{equation}
where $U_{tot}(\Delta t)=U_S(\Delta t)U_I(\Delta t)U_E(\Delta t)$, and:
\begin{equation}
\label{eqn:operators}
    U_S(\Delta t)=\exp(-i g_S H_S \Delta t),\quad U_I(\Delta t)=\exp(-i g_I H_I \Delta t),\quad U_E(\Delta t)=\exp(-i g_E H_E \Delta t).
\end{equation}
Note that we are keeping the same notation as in Section~\ref{sec:collModelSingle}, where $U_S$ describes a free system evolution (not shown in Fig.~\ref{fig:entangled}), while $U_I$ is the operator for the ancillas-subsystems interaction. Moreover, we have added the environment operator $U_E$, driven by the Hamiltonian $H_E$ with magnitude $g_E$, that entangles the ancillas.  

Our aim is to derive a GKLS master equation generated by the quantum map in Eq.~\eqref{eqn:singleTimeEnt} in the standard limit of an infinitesimal timestep. To do so, we assume to be in the regime of \textit{fast environment Hamiltonian}\footnote{We show in Appendix~\ref{sec:slow} the issues that arise if one does not assume to be in this regime.}, that is:
\begin{equation}
    \label{eqn:fastEnvHam}
    \lim_{\Delta t\rightarrow 0^+} g_E \Delta t = \mu,
\end{equation}
where $\mu$ is a finite dimensionless constant. Furthermore, we also assume the standard limits for $g_S$ and $g_I$ that were employed in Section~\ref{sec:collModelSingle}. Specifically, $\lim_{\Delta t\rightarrow 0^+}g_I^2\Delta t=\gamma$, where $\gamma$ is a constant with the units of the inverse of time, and $\lim_{\Delta t\rightarrow 0^+}g_S^2\Delta t=0$. Note that this implies $g_S\ll g_I\ll g_E$, which clarifies the nomenclature ``fast environment Hamiltonian''. Therefore, using Eq.~\eqref{eqn:fastEnvHam} and assuming that $\Delta t$ is very small, we can write $U_E(\Delta t)\rightarrow U_E(\mu)=\exp(-i \mu H_E)$, and rewrite Eq.~\eqref{eqn:singleTimeEnt} as:
\begin{equation}
    \label{eqn:singleTimeEntEnt}
    \phi_{\Delta t}^{(ent)}[\rho_S]=\Tr_E[U_S(\Delta t)U_I(\Delta t)\rho_S\otimes\rho_E(\mu)U_I^\dagger(\Delta t)U_S^\dagger(\Delta t)],
\end{equation}
where we have introduced the entangled state of the ancillas:
\begin{equation}
\label{eqn:entStatPure}
    \rho_E(\mu)=U_E(\mu)\rho_E(0)U_E^\dagger(\mu).
\end{equation}
Note that, contrary to the decomposition of the operator in Eq.~\eqref{eqn:opDecomposedSI}, we cannot commute $U_E$ with $U_I$ or $U_S$, as $\mu$ is not an infinitesimal parameter. Therefore, $U_E(\mu)$ must always be the first operator that is implemented during a single timestep. This is reasonable, as we need to prepare the ancillas in an entangled state \textit{before} starting the multipartite collisions.

Finally, we point out that we may also choose a mixed entangled state of the ancillas by a linear combination of different states written as in Eq.~\eqref{eqn:entStatPure}. On a quantum computer, this may be realized either by a purification of the mixed state in an enlarged Hilbert space through additional ancillas, or by random sampling the preparation of different entangled pure states written as in  Eq.~\eqref{eqn:entStatPure}. The most general state we can employ in the collision model can then be written as:
\begin{equation}
    \label{eqn:entStatMixed}
    \rho_E^{(ent)}=\sum_k p_k\rho_{E}^{(k)}(\mu_k),\text{ with } p_k\geq 0 \text{ and }\sum_k p_k=1,
\end{equation}
where each $\rho_E^{(k)}$ is given by Eq.~\eqref{eqn:entStatPure} with a different environment Hamiltonian and a different $\mu_k$.

\subsection{Derivation of the master equation and its properties}
We can derive a GKLS master equation starting from Eq.~\eqref{eqn:singleTimeEntEnt} following the same procedure as in Section~\ref{sec:collModelSingle}, with an interaction Hamiltonian that can be decomposed as a sum of local ancilla-subsystem terms:
\begin{equation}
    \label{eqn:intHamEnt}
    H_I=\sum_{m=1}^M \sum_\alpha F_{m,\alpha}\otimes B_{m,\alpha},
\end{equation}
where $F_{m,\alpha}$ and $B_{m,\alpha}$ are respectively local system and ancillary operators. 

Using Eqs.~\eqref{eqn:collisionDer2} and~\eqref{eqn:collisionGKLS}, in the limit of infinitesimal timestep we obtain:
\begin{equation}
\label{eqn:derEnt1}
    \frac{d\rho_S}{dt}=\mathcal{L}_{ent}[\rho_S]=-i g_S[H_S,\rho_S]+\gamma \Tr_E[\mathcal{D}_{H_I}[\rho_S\otimes\rho_E^{(ent)}]],
\end{equation}
where $\mathcal{D}_A[\rho]=A \rho A -\frac{1}{2}\{\rho,A^2\}$, and we have assumed\footnote{Note that this assumption may not be trivial, contrary to the scenarios with separable ancillas (e.g., with $\sigma_{E_j}^+$ as ancilla operator, which has zero mean value on thermal ancillary states), and care must be taken to choose a suitable $H_I$.} $\Tr_E[H_I\rho_S\otimes\rho_E^{(ent)}]=0$. Inserting Eq.~\eqref{eqn:intHamEnt} into Eq.~\eqref{eqn:derEnt1}, we find:
\begin{equation}
\label{eqn:gklsEnt}
    \mathcal{L}_{ent}[\rho_S]=-i g_S[H_S,\rho_S] +\sum_{m,m'=1}^M \sum_{\alpha,\alpha'}\tilde{\gamma}_{m,\alpha,m',\alpha'}\left(F_{m,\alpha}\rho_S F_{m',\alpha'}-\frac{1}{2}\{F_{m',\alpha'}F_{m,\alpha},\rho_S\} \right),
\end{equation}
where the coefficients of this multipartite master equation are given by \footnote{Note that, contrary to Eq.~\eqref{eqn:gklsMultipartite}, we are not inserting a dagger in Eq.~\eqref{eqn:gklsEnt}, so these coefficients refer to pairs of GKS operators that are different from the ones in Eq.~\eqref{eqn:gklsMultipartite}.}:
\begin{equation}
\label{eqn:coeffEnt}
    \tilde{\gamma}_{m,\alpha,m',\alpha'}=\gamma \Tr_E[B_{m',\alpha'}B_{m,\alpha}\rho_E^{(ent)}].
\end{equation}
Note that the non-locality of Eq.~\eqref{eqn:gklsEnt} is due to the fact that the autocorrelation functions in Eq.~\eqref{eqn:coeffEnt} are evaluated on an entangled state of the environment.

Eq.~\eqref{eqn:gklsEnt} resembles the GKLS master equation for multipartite systems we introduced in Eq.~\eqref{eqn:gklsMultipartite}. Indeed, in principle we are able to encode any possible GKS operator $F_{m,\alpha}$ in the interaction Hamiltonian in Eq.~\eqref{eqn:intHamEnt}. However, we immediately find a limitation of Eq.~\eqref{eqn:gklsEnt} that is due to the structure of its coefficient, and that inevitably limits its generality. In particular, we observe that $[B_{m,\alpha},B_{m',\alpha'}]=0$ if $m\neq m'$, because they are operators acting on different ancillas. As an immediate consequence, we find:
\begin{equation}
\label{eqn:symmetryEntCoeff}
 \tilde{\gamma}_{m,\alpha,m',\alpha'}=\tilde{\gamma}_{m',\alpha',m,\alpha}.
\end{equation}
This is a serious limitation. For instance, suppose that the system is composed of two qubits ($M=2$) and $F_{m,1}=\sigma_m^-$, $F_{m,2}=\sigma_m^+$. Then, Eq.~\eqref{eqn:symmetryEntCoeff} implies that, in the final GKLS master equation, the term $\sigma_1^-\rho_S \sigma_2^+$ will have the same coefficient as the term $\sigma_2^+\rho_S\sigma_1^-$, and, as a consequence of the properties of the master equation, these coefficients have the same module as the ones of the terms  $\sigma_1^+\rho_S \sigma_2^-$ and $\sigma_2^-\rho_S\sigma_1^+$. This means that Eq.~\eqref{eqn:gklsEnt} is not able, for instance, to reproduce the action of a master equation for a global thermal bath at \textit{finite} temperature (e.g., see Ref.~\cite{Cattaneo2019b}), as an equal decay rate for the emission and for the absorption in the cross terms is a signature of \textit{infinite} global temperature.

Possible generalizations of Eq.~\eqref{eqn:gklsEnt}, with the aim of tuning the coefficients $\tilde{\gamma}$ in the most general way, may be implemented by assigning two independent entangled ancillas to each possible pair of subsystems. Each ancilla would then collide locally with the corresponding subsystem. Each pair of collisions would generate a term of the GKLS master equation that couples two subsystems with tunable coefficients given by Eq.~\eqref{eqn:coeffEnt}. Furthermore, more freedom may be granted by adding some tunable weights in the subsystem-ancilla interactions in Eq.~\eqref{eqn:intHamEnt}. The number of necessary ancillas to achieve this generality is clearly exponential in the number of subsystems. Moreover, note that the constraint given by Eq.~\eqref{eqn:symmetryEntCoeff} is always valid for this type of collision model.

We have finally understood that the collision model with entangled ancillas gives rise to a limited variety of multipartite GKLS equations, and does not provide the generality that is granted by the multipartite collision model. However, quite intuitively, Eq.~\eqref{eqn:gklsEnt} is particularly convenient to simulate the Markovian dynamics that is microscopically generated by entangled separate baths \cite{breuer2002theory,Cattaneo2019b}. Indeed, the structure of the coefficients of this type of master equation  is quite similar  to the one in Eq.~\eqref{eqn:coeffEnt}. We will provide an example of such a master equation in the next subsection. 
 
\subsection{Example: two-mode squeezed ancillas}
\label{sec:exa}
 We provide here the example of a master equation generated by the collision model with entangled ancillas when the system is made of two qubits ($M=2$) and the ancillas are two bosonic modes prepared in a two-mode squeezed thermal state. If $S(\zeta)=\exp\left(\zeta b_1^\dagger b_2^\dagger-\zeta^* b_1b_2\right)$, where $b_1$ and $b_2$ are the bosonic annihilation operator of respectively the first and second ancillary mode, and $\zeta$ is a complex variable, we have \cite{ferraro2005gaussian}:
\begin{equation}
\rho_E^{(ent)}=S(\zeta) \rho_{th}(N_1)\otimes\rho_{th}(N_2)S(\zeta)^\dagger,
\end{equation}
where $\rho_{th}(N_j)$ is a thermal state with average photon number equal to $N_j$. Let us suppose that the subsystems-ancillas interaction Hamiltonian reads:
\begin{equation}
H_I=\sigma_1^-b_1^\dagger+\sigma_2^-b_2^\dagger+H.c.
\end{equation}
Then, the coefficients of the master equation (Eq.~\eqref{eqn:coeffEnt}) are given by $\tilde{\gamma}_{m,\alpha,m',\alpha'}=\gamma\Tr_E[b_{m'}^{(\alpha')}b_m^{(\alpha)}\rho_E^{(ent)}]$, where the labels $\alpha,\alpha'=\dagger,\cdot$ indicate whether or not the bath operators are daggered. By using the cyclic property of the trace, we obtain: 
\begin{equation}
\tilde{\gamma}_{m,\alpha,m',\alpha'}=\gamma\Tr_E[S(\zeta)^\dagger b_{m'}^{(\alpha')}S(\zeta) S(\zeta)^\dagger b_m^{(\alpha)}S(\zeta)\rho_{th}(N_1)\otimes\rho_{th}(N_2)].
\end{equation}
According to the Baker-Campbell-Hausdorff formula, we have:
\begin{equation}
\label{eqn:squeezOp}
\begin{split}
&S(\zeta)^\dagger b_1 S(\zeta)= \cosh r \,b_1+e^{i\psi}\sinh r\, b_2^\dagger\\
&S(\zeta)^\dagger b_2 S(\zeta)=\cosh r\, b_2+e^{i\psi}\sinh r\, b_1^\dagger,
\end{split}
\end{equation}
where $\zeta= r e^{i\psi}$. Using the above relations, we observe that $\Tr_E[b_j b_k^\dagger\rho_E^{(ent)}]=0$ for all $j\neq k$, $\Tr_E[b_j^2\rho_E^{(ent)}]=0$ for all $j$. Moreover, using Eq.~\eqref{eqn:squeezOp} we immediately observe that $\Tr_E[b_j \rho_E^{(ent)}]=0$, therefore the condition for the derivation of Eq.~\eqref{eqn:derEnt1} is fulfilled. 

Finally, we use the above results to derive the master equation for the collision model with entangled ancillas, given by Eq.~\eqref{eqn:gklsEnt} (for simplicity, we ignore the free system dynamics driven by $H_S$):
\begin{equation}
\label{eqn:gklsEx}
\begin{split}
\mathcal{L}^{(ent)}[\rho_S]=&\gamma_1^\downarrow(\sigma_1^-\rho_S\sigma_1^+-\frac{1}{2}\{\sigma_1^+\sigma_1^-,\rho_S\})+\gamma_1^\uparrow(\sigma_1^+\rho_S\sigma_1^--\frac{1}{2}\{\sigma_1^-\sigma_1^+,\rho_S\})\\
&+\gamma_2^\downarrow(\sigma_2^-\rho_S\sigma_2^+-\frac{1}{2}\{\sigma_2^+\sigma_2^-,\rho_S\})+\gamma_2^\uparrow(\sigma_2^+\rho_S\sigma_2^--\frac{1}{2}\{\sigma_2^-\sigma_2^+,\rho_S\})\\
&+\gamma_c(\sigma_1^-\rho_S\sigma_2^--\frac{1}{2}\{\sigma_2^-\sigma_1^-,\rho_S\}+\sigma_2^-\rho_S\sigma_1^--\frac{1}{2}\{\sigma_1^-\sigma_2^-,\rho_S\})\\
&+\gamma_c^*(\sigma_1^+\rho_S\sigma_2^+-\frac{1}{2}\{\sigma_2^+\sigma_1^+,\rho_S\}+\sigma_2^+\rho_S\sigma_1^+-\frac{1}{2}\{\sigma_1^+\sigma_2^+,\rho_S\}).\\
\end{split}
\end{equation} 
The coefficients read:
\begin{equation}
\begin{split}
&\gamma_j^\downarrow=\gamma\left(\cosh^2 r \,(N_j+1)+\sinh^2 r N_{j+1}\right),\qquad \gamma_j^\uparrow=\gamma\left(\cosh^2 r \,N_j +\sinh^2 r (N_{j+1}+1)\right),\\
&\gamma_c=\gamma\cosh r \,\sinh r \,e^{i\psi}(N_1+N_2+1),\\
\end{split}
\end{equation}
where for simplicity we have used a notation such that $N_3=N_1$.
Note that the coefficient associated with the term $\sigma_1^-\rho_S\sigma_2^-$ must be equal to the one for $\sigma_2^-\rho_S\sigma_1^-$, according to Eq.~\eqref{eqn:symmetryEntCoeff}, and analogously for all the other cross terms.

A microscopic model of an open quantum system whose master equation would resemble Eq.~\eqref{eqn:gklsEx} may be two qubits interacting with two local, separate environments \cite{Cattaneo2019b}, where the initial state of the baths is such that the modes of each environment with the same frequency are entangled in two-mode squeezed thermal states. The same structure of the coefficients would then be recovered in the singular-coupling limit \cite{breuer2002theory}, corresponding to autocorrelation functions of the environments that decay instantaneously. 

A master equation with the same structure as Eq.~\eqref{eqn:gklsEx} can be simulated on a quantum computer by using ancillary qubits prepared in the entangled state $b_{ee}\ket{ee}+b_{gg}\ket{gg}$, where $\ket{e}$ and $\ket{g}$ are respectively the qubit excited and ground state, while $b_{ee}$ and $b_{gg}$ are normalization coefficients that we can freely choose \cite{Daryanoosh2018}. To properly tune the values of the coefficients of the master equation, one may then add some weights in the interaction Hamiltonian Eq.~\eqref{eqn:intHamEnt}, as done for the MCM in Eq.~\eqref{eqn:ElColMCM}, and/or apply more collisions with new ancillas prepared in a different entangled state.

\section{Conclusions}
\label{sec:conclusions}
In this paper, we have analyzed four different collision models for multipartite open quantum systems that reproduce a GKLS master equation in the limit of an infinitesimal timestep. First, in Section~\ref{sec:collModelSingle} we have reviewed the derivation of such a master equation for the standard collision model for a single open system. Then, we have discussed the ideas and limitations of the collision models for multipartite open dynamics in Section~\ref{sec:collModels}, also showing how to simulate them on a quantum computer for some very simple scenarios. We will briefly summarize these discussions here.

The \textit{multipartite collision model} \cite{Cattaneo2021} (MCM) is the most general protocol we have discussed in this work. It can reproduce the Markovian master equation of any multipartite open quantum systems thanks to the structure of its collisions, which are based on second-order Suzuki-Trotter decomposition. To achieve this level of generality, however, it requires a large number of ancillas, which can be reduced in the presence of symmetries in the master equation.

The \textit{cascade collision model} \cite{Giovannetti2012,Giovannetti2012a} can simulate master equations for cascade dynamics, where the environment ancillas collide with the subsystems in a time-ordered way (i.e., first with the first subsystems, then with the second one, and so on). However, it cannot efficiently reproduce any open dynamics that does not have this cascade structure.

The \textit{composite collision model} \cite{Lorenzo2017} is a very simple and elegant collision model that simulates local master equations for multipartite systems, where each ancilla (say one for each subsystem) collides locally with a single subsystem only, while the different subsystems interact only through the free system dynamics driven by the Hamiltonian $H_S$. However, it cannot simulate global master equations, that, for instance, display collective effects.

The \textit{collision model with entangled ancillas} \cite{Daryanoosh2018} can naturally simulate master equations for correlated environments (e.g., see the example we proposed in Section~\ref{sec:exa}). However, we have shown in Section~\ref{sec:properties} that the coefficients of the master equations it generates are constrained by the condition expressed in Eq.~\eqref{eqn:symmetryEntCoeff}. This limits the range of master equations it can simulate and, for instance, it is not suitable to reproduce the dynamics driven by a global bath at finite temperature.

\section*{Acknowledgements}
We would like to thank Gabriele De Chiara for interesting discussions on collision models for global master equations. \mar{M.C. would also like to thank Sergey Filippov for some useful comments on the structure of Eq.~\eqref{eqn:cascadeGKLS} and on the example discussed in Appendix~\ref{sec:exDiff}.} M.C.   and   S.M.   acknowledge   financial support from the Academy of Finland via the Centre of Excellence  program  (Project  No.   336810  and  Project No.   336814). M.C., G.L.G. and  R.Z.  acknowledge  financial  support  from  Centers and  Units  of  Excellence  in  R\&D  (MDM-2017-  0711) and from MICINN/AEI/FEDER and CAIB for projects PID2019-109094GB-C21/AEI/10.13039/501100011033 and PRD2018/47. G.L.G. is funded by the Spanish MEFP/MiU and co-funded by the University of the Balearic Islands through the Beatriz Galindo program (BG20/00085).

\appendix
\section{\mar{An example on the difference between the master equations generated by the MCM and the cascade collision model}}
\label{sec:exDiff}
In Section~\ref{sec:cascade} we have presented the cascade collision model and the master equation it generates, given by Eq.~\eqref{eqn:cascadeGKLS}. We have shown that its generality is limited by the structure of the global dissipator in Eq.~\eqref{eqn:globalDissipator1}, and that the difference between this master equation and the one generated by the MCM (Section~\ref{sec:MCM}) is due to the time order of the collisions, and can be captured by the Lamb-shift term given by Eq.~\eqref{eqn:lambShift}. In this appendix, we will shed light on this subtle point by providing a very simple example based on these master equations.

Let us consider a system made of two subsystems only, say subsystem $1$ and subsystem $2$, and a single ancillary qubit that we label as $E$. For each subsystem, we aim to reproduce the action of only a single GKS operator $F_j$ in the master equation, where $j=1,2$ is labeling the subsystems. Moreover, the ancilla is initialized in the ground state before the collision. For simplicity, we will not implement the free system evolution. We create the elementary-collision operators as:
\begin{equation}
    U_j(\Delta t)=\exp(-i g_I H_j\Delta t),\text{ with } H_j=\lambda_j F_j\sigma_E^++\lambda_j^*F_j^\dagger \sigma_E^-,\quad j=1,2.
\end{equation}
We will employ this collision operator for both the cascade collision model and the MCM. In contrast, the total operator for a single timestep of each collision model will be:
\begin{equation}
    U_{cas}(\Delta t)=U_2(\Delta t)U_1(\Delta t),\quad U_{\text{MCM}}(\Delta t)=U_2(\Delta t/2)U_1(\Delta t)U_2(\Delta t/2),
\end{equation}
respectively for the cascade and multipartite collision model. The action of the quantum map reproducing a single timestep of, for instance, the MCM is then
\begin{equation}
    \phi_{\text{MCM}}[\rho_S]=\Tr_E[U_{\text{MCM}}(\Delta t)\rho_S\otimes\ket{0}_E\!\bra{0}U_{\text{MCM}}^\dagger(\Delta t)],
\end{equation}
and analogously for the cascade model. In the standard limit of infinitesimal timestep, with $\Delta t\rightarrow 0^+$ and $g_I^2\Delta t\rightarrow \gamma$, we can then expand each elementary collision up to the second order in $g_I\Delta t$:
\begin{equation}
    U_j(\Delta t)=\mathbb{I}_{SE}-i g_I\Delta t H_j-\frac{g_I^2\Delta t^2}{2}H_j^2+O(g_I^3\Delta t^3).
\end{equation}
Let us now focus on the cross terms between the subsystems that appear in the Liouvillian $\mathcal{L}_{\text{MCM}}=\lim_{\Delta t\rightarrow 0^+}\frac{\phi_{\text{MCM}}-\mathbb{I}_S}{\Delta t}$, and analogously for the cascade model. Due to the second-order Suzuki-Trotter decomposition in Eq.~\eqref{eqn:secondorderSuzuki}, $U_{\text{MCM}}(\Delta t)$ is equivalent to $\exp(-i g_I (H_1+H_2)\Delta t)$ up to a negligible remainder, and it is straightforward to see that the cross terms in the Liouvillian applied to $\rho_S$ will be (up to some constants):
i) $\Tr_E[H_1\rho_S\otimes\ket{0}_E\!\bra{0} H_2]$, ii) $\Tr_E[H_2\rho_S\otimes\ket{0}_E\!\bra{0} H_1]$, iii) $\Tr_E[H_2H_1\rho_S\otimes\ket{0}_E\!\bra{0}]$,
iv) $\Tr_E[\rho_S\otimes\ket{0}_E\!\bra{0}H_1H_2]$, v) $\Tr_E[\rho_S\otimes\ket{0}_E\!\bra{0}H_2H_1]$, vi) $\Tr_E[H_1H_2\rho_S\otimes\ket{0}_E\!\bra{0}]$. The symmetry in the structure of the GKS operators between subsystem $1$ and subsystem $2$ is evident. After some trivial algebra, we obtain the final Liouvillian:
\begin{equation}
\label{eqn:liouvillianMCM}
\begin{split}
\mathcal{L}_{\text{MCM}}[\rho_S]=\sum_{j,k=1,2}\gamma \lambda_j\lambda_k^* \left(F_j\rho_S F_k^\dagger-\frac{1}{2}\{F_k^\dagger F_j,\rho_S\}\right).
\end{split}
\end{equation}

Let us now analyze the cascade collision model. Due to the order of the elementary collisions in a single timestep, we immediately realize that the cross terms in the cascade Liouvillian will be (up to some constants) the same as i), ii), iii) and iv) for the MCM. However, v) and vi) are missing, because the ancilla first collides with subsystem $1$ and then with subsystem $2$, without bouncing back towards subsystem $1$. Therefore, the symmetry between the two subsystems is broken. We thus understand that the second-order Suzuki-Trotter decomposition is \textit{the way to ensure that the subsystems in the expansion of the collision operators in the quantum map are treated on the same footing}, i.e., all the possible combinations of GKS operators are present in the Liouvillian. This is also a requirement to obtain the most general GKLS equation for multipartite systems, expressed by Eq.~\eqref{eqn:gklsMultipartite}. In contrast, the Liouvillian of the cascade collision model does not have two crucial cross terms:
\begin{equation}
\label{eqn:liouvillianCascade}
\begin{split}
\mathcal{L}_{cas}[\rho_S]=\sum_{j=1,2}\gamma \abs{\lambda_j}^2 \left(F_j\rho_S F_j^\dagger-\frac{1}{2}\{F_j^\dagger F_j,\rho_S\}\right)+\gamma\left[\lambda_1\lambda_2^* (F_1\rho_S F_2^\dagger -F_2^\dagger F_1\rho_S)+H.c.\right].
\end{split}
\end{equation}
The absence of the terms $\rho_S F_2^\dagger F_1$ and $ F_1^\dagger F_2 \rho_S$ has important consequences on the dynamics of the two subsystems, which reflect the time order with which the ancilla collides against them. For instance, Eq.~\eqref{eqn:liouvillianCascade} for two bosonic modes (specifically, with $F_j=a_j^\dagger$) is exactly the master equation derived by Bergou and Hillery for two masers \cite{Bergou1991}, which displays an unbalanced dynamics between the two cavities. For instance, if the initial state of the maser fields is the ground, the authors found that the photon number in the second cavity grows faster than in the first.

Following Ref.~\cite{Cusumano2017a} and our discussion in Section~\ref{sec:cascade}, the cascade Liouvillian in Eq.~\eqref{eqn:liouvillianCascade} can be brought into the symmetric structure of Eq.~\eqref{eqn:liouvillianMCM}, but then a Lamb-shift term appears in the master equation, which restores the time order of the collisions. This Lamb-shift term reads:
\begin{equation}
    \label{eqn:lambShiftEx}
    H_{LS}=\frac{\lambda_1\lambda_2^* F_1F_2^\dagger-\lambda_1^*\lambda_2 F_1^\dagger F_2}{2i}.
\end{equation}

\section{Slow environment Hamiltonian}
\label{sec:slow}
In this appendix, we quickly sketch the idea of why we need to assume the regime of fast environment Hamiltonian (see Eq.~\eqref{eqn:fastEnvHam}) for the derivation of the master equation of the collision model with entangled ancillas discussed in Section~\ref{sec:properties}. First, let us assume the condition for the \textit{slow environment Hamiltonian} on the magnitude $g_E$ introduced in Eq.~\eqref{eqn:operators}, that is:
\begin{equation}
    \label{eqn:slowEnvH}
    \lim_{\Delta t\rightarrow 0^+} g_E \Delta t=0.
\end{equation}
Our aim is to try to derive a GKLS master equation similar to Eq.~\eqref{eqn:gklsEnt} in the regime of slow environment Hamiltonian and following the idea of the standard derivation of a collision model in Section~\ref{sec:collModelSingle}. Quite intuitively, we need to require $g_S\ll g_I \ll g_E$, since if the magnitude of the environment Hamiltonian is not much larger than the magnitude of the interaction Hamiltonian, then the collision will happen before the ancillas get entangled, and no cross terms will arise in the master equation. 

The reader can verify that, by expanding the operator $U_{tot}(\Delta t)$ introduced in Eq.~\eqref{eqn:singleTimeEnt} up to the fourth order in $g_j\Delta t\ll 0$ ($j=S,I,E$), only some terms proportional to powers of $g_I$ and $g_E$ cannot be trivially removed following the same lines as for the derivation in Section~\ref{sec:collModelSingle} (the only term proportional to $g_S$ that we keep in the master equation is of the first order and generates the free system dynamics, as usual). For simplicity, we won't explicitly write  these terms here.

A term of the first order in $g_I\Delta t$ (plus additional orders in $g_E\Delta t$) can be removed by requiring $\lim_{\Delta t\rightarrow 0^+}\Tr_E[[H_I,\rho_S\otimes U_E(\Delta t)\rho_E(0)U_E^\dagger(\Delta t)]]=0$. This assumption is reasonable \cite{Daryanoosh2018}, and it is basically equivalent to the assumption we make to derive Eq.~\eqref{eqn:gklsEnt} in the regime of fast environment Hamiltonian. Then, after dividing by $\Delta t$ to obtain the master equation (see Eq.~\eqref{eqn:collisionGKLS}), we are left with a term proportional to $g_I^2\Delta t$ (the same term as in Section~\ref{sec:collModelSingle}), another one proportional to $g_I^2g_E\Delta t^2$, and the last one that is proportional to $g_I^2g_E^2\Delta t^3$. To obtain a global master equation for the multipartite system, we need to keep at least one of the last two terms, which contain the entangling environment Hamiltonian that is creating the correlations between the subsystems. However, due to the requirement $g_E\Delta t\ll 1$, the term proportional to $g_I^2 \Delta t$ is dominating over the other two terms. Therefore, the uncorrelated term is the only one that is relevant to the dynamics we are describing here, while the ancilla-driven correlations must be very small. In other words, it is not possible to derive a consistent GKLS master equation for the collision model with entangled ancillas outside the regime of fast environment Hamiltonian. 

\bibliographystyle{apsrev4-1}
\bibliography{biblio}

\end{document}